\documentclass[runningheads]{llncs}
\usepackage[T1]{fontenc}

\usepackage{subcaption,graphicx,float}
\usepackage{amsmath}
\usepackage{booktabs}
\usepackage{hyperref}

\usepackage{xcolor}

\usepackage[english]{babel}
\usepackage[protrusion=true,expansion=true]{microtype}

\usepackage{textcomp}

\usepackage[
  backend=biber,
  style=lncs,
  maxbibnames=8,
  minbibnames=8
]{biblatex}
\begin{document}

\title{Towards In Silico Cancer Therapy Design: An Agent-Based Approach for GPU-Accelerated Molecular Pathway Simulation\thanks{A preliminary version of this work appeared in the Collections of Short Papers of CIBB 2025 (20th International Conference on Computational Intelligence Methods for Bioinformatics and Biostatistics, Milan, 10--12 September 2025).}}
\author{Stefano Maestri\inst{*,1}\orcidID{0000-0002-7615-4060}}
\authorrunning{Stefano Maestri}
\titlerunning{Towards In Silico Cancer Therapy Design}
\institute{University of Camerino, Camerino, 62032, Italy\\
\email{stefano.maestri@acm.org}}
\maketitle
\begin{abstract}
Agent-based modelling is gaining recognition as a powerful approach for simulating complex cellular pathways, owing to its ability to reproduce emergent biological behaviours without requiring extensive kinetic parameterisation. In this article, we present a GPU-accelerated agent-based simulator specifically designed to model and analyse signalling pathways involved in cancer progression, and to evaluate therapeutic interventions. Our approach leverages the computing capabilities of FLAME GPU 2, a GPU-accelerated agent-based modelling framework, to efficiently manage simulations involving millions of molecules interacting within a three-dimensional environment. Each molecule is represented as an autonomous agent with defined physical properties, capable of binding, releasing reaction products, migrating between compartments, and interacting based on spatial proximity. An intuitive graphical interface supports model construction, parameter setup, and real-time modification of treatment strategies. As the primary focus of this paper, we validate the simulator on the MAPK/ERK cascade affected by the BRAFV600E mutation, demonstrating that it accurately reproduces dose--response trends observed in clinical data and outperforms both deterministic models and our prior agent-based implementations. 
A second case study extends the approach to nuclear signalling by reproducing the dynamics of cFos expression and phosphorylation. This demonstrates the simulator’s ability to capture compartmentalised regulation, reproducing transient mRNA responses and protein accumulation, including the effect of an unresolved negative transcriptional regulator. Together, these results show that GPU-accelerated ABM can faithfully replicate both drug response and emergent gene expression dynamics, providing a scalable and biologically grounded computational tool for supporting precision oncology.

\end{abstract}

\newpage
\section{Introduction}
\label{sec:introduction}

Cancer therapy has witnessed significant advancements through computational modelling techniques, with agent-based models (ABMs) increasingly recognised for their potential to simulate complex biological systems by modelling individual entities (agents) and their interactions~\cite{an_agent-based_2009}. ABMs offer unique advantages over traditional deterministic or kinetic models, notably their ability to reproduce emergent biological behaviours based on simple local-interaction rules, minimising dependency on extensive kinetic data, which are often difficult to obtain experimentally~\cite{belenchia_agent-based_2021,leighow_agent-based_2022,maestri_agent-based_2022}.

Our previous study, presented at CIBB 2024, introduced an agent-based simulator implemented on Repast4Py, a recently released High-Performance Computing (HPC) framework, and investigated the inhibitory effects of the targeted therapy drug dabrafenib on the BRAFV600E-MEK-ERK signalling cascade, an aberrant pathway particularly relevant in melanoma progression~\cite{maestri_cutting_2025}. To manage computational complexity within the constraints of mid-range hardware, this initial implementation relied on a spatial abstraction method, referred to as the \textit{slicing} of the simulation volume. This method effectively reduced computational demand by simulating molecular interactions within a thin slice of the three-dimensional space. Despite successfully validating the approach against clinical data, this spatial simplification introduced limitations, particularly regarding the accurate representation of low molecular concentrations, potentially affecting the fidelity of simulation outcomes.

To overcome these obstacles, we moved away from the Repast4Py implementation, whose performance scales effectively only in distributed environments, and sought a solution capable of delivering a higher throughput on a single machine. Exploiting the massive parallelism of modern Graphics Processing Units (GPUs) proved ideal, and the rapidly maturing FLAME GPU 2 agent-based modelling framework (which reached a stable release candidate at the time of our study) offered exactly the capabilities and performance we needed~\cite{richmond_flame_2023}. 

Hence, the present work introduces a GPU-accelerated agent-based simulator built on FLAME GPU 2 (v2.0.0-rc.2), which is capable of handling realistic cellular volumes without any spatial abstraction. This tool significantly enhances scalability and computational efficiency by exploiting GPU parallelism, thus enabling simulations involving millions of molecules interacting within fully three-dimensional environments. Eliminating the need for volume slicing allows for a more precise representation of biological phenomena, particularly those sensitive to accurate spatial distribution and concentration gradients.

In this article, we demonstrate the simulator's new capabilities by revisiting the dabrafenib-mediated inhibition of the MAPK/ERK pathway in melanoma cells harbouring the BRAFV600E mutation~\cite{hauschild_dabrafenib_2012}. Building upon our prior work, we further validate the simulator by replicating clinical dose--response data with higher accuracy.
Additionally, we introduce an intuitive graphical interface that facilitates user interactions with the simulation, enabling real-time adjustments of therapeutic strategies and automatic retrieval of essential molecular parameters from biochemical databases. These advancements refine the predictive accuracy of our simulations and enhance usability, positioning the new simulator as a reliable in silico support for precision oncology.

Beyond this pharmacological setting, we also tested the simulator in a second, mechanistically distinct case study: the regulation of cFos downstream of MAPK/ERK activation, as characterised by Nakakuki et al.~\cite{nakakuki_ligand-specific_2010}. 
This case focuses on transcriptional dynamics and nucleocytoplasmic compartmentalisation, evaluating the model’s ability to reproduce transient mRNA responses, sustained protein accumulation, and the regulatory role of an unresolved transcriptional repressor. 

Together, the two case studies offer a rigorous benchmark for the framework, demonstrating that GPU-accelerated ABM can reproduce both targeted drug responses in a clinical setting and emergent gene expression dynamics in complex signalling systems.

\section{Materials and Methods}
\label{sec:materials_and_methods}

The simulator described in this work is built on FLAME GPU 2, a framework specifically designed for developing agent-based models simulated on GPU architectures~\cite{richmond_flame_2023}. FLAME GPU 2 provides a flexible environment for defining agent interactions based on spatial proximity and supports rapid simulation execution by leveraging NVIDIA's CUDA technology~\cite{nickolls_scalable_2008}. Our ABM follows a hybrid implementation with a host-side control code written in Python and GPU-accelerated agent functions implemented in CUDA C++.

For detailed information on the physical and biochemical principles embedded within our agent behaviours---including diffusion, molecular interaction logic, and reaction kinetics---readers are referred to Maestri et al.~\cite{maestri_agent-based_2022} and our previous article presented at CIBB 2024~\cite{maestri_cutting_2025}. Briefly, agents represent molecules as spheres whose radii are derived from molecular weights, with diffusion coefficients calculated using the Stokes-Einstein equation. Enzymatic reaction timing (product formation and release) is based on their turnover numbers ($k_{\mathrm{cat}}$).
These parameters collectively ensure realistic spatial and temporal behaviour within the simulated cellular environment.

The uniqueness of our agent-based approach lies in requiring a limited amount of experimental data, often readily available, for the initial setup\footnote{The key parameters are simulation volume, initial concentrations, molecular weights, and enzymatic turnover numbers, the latter two often available from databases such as UniProt and BRENDA. Modelling compartments requires providing compartment volumes and translocation rates; if gene expression is also taken into account, expression rates must be specified---these additional values are generally harder to obtain and may require estimation.}. In particular, unlike other approaches, such as models based on differential equations, \textit{our simulations do not require the kinetic parameters needed to reproduce the formation of molecular complexes}, as these properties emerge from the local interactions of the agents.

Our GPU-based simulator offers an integrated graphical interface that facilitates the creation and management of simulation models. Alternatively, models can be defined as a structured JSON file, which is loaded as input for the simulations.
The model-building process primarily involves defining the types of molecular species involved. Our ABM paradigm categorises molecules into four functional classes:
\begin{itemize}
    \item \textbf{Active} - molecules capable of binding to other molecules in the system. These are typically enzymes, such as kinases or phosphatases, that activate or inhibit a specific substrate.
    \item \textbf{Inactive} - molecules that act as substrates for active species.    
    \item \textbf{Inhibited} - molecules that have already bound to an inhibitor, such as a drug.    
    \item \textbf{Expressors} - molecules without a specific binding target, which continuously produce new molecules at a fixed rate determined by experimental data\footnote{The author acknowledges that, in experimental genetics and oncology, the term ``expressor'' is commonly used as shorthand for a biological system (cell line, clone, or tumour) classified by the presence or level of a specific marker (e.g., ``high-expressor'' cell lines). Here, the term is intentionally generalised to denote an abstract agent class that continuously emits new molecules at a defined rate within the simulation framework.}.   
\end{itemize}

Active and inactive molecules interact during the simulation, while expressors are responsible for producing new molecules in models that include gene activation or repression dynamics. Only active, inactive, and expressor species must be defined during the model setup. Inhibited molecules emerge during the simulation as a result of processes like drug action, and therefore do not require explicit specification. 

For each molecular species, the user can provide properties such as molecular weight, initial concentration, and localisation (cytosolic or nuclear), along with the reactions it participates in, including interacting partners, rates, and resulting products. Turnover numbers and molecular weights can be automatically retrieved from publicly available databases, specifically BRENDA and UniProt, to streamline model setup and ensure consistency with experimentally validated values.

A dedicated consistency checking feature validates cross-references among molecules before execution. It verifies the presence of reciprocal interaction definitions, the existence of all referenced molecules, the plausibility of molecular weights, and the compatibility between volume settings and molecular concentrations.

The simulator enables users to designate specific molecular species for continuous monitoring, allowing them to define concentration thresholds within specified time windows. For example, by selecting phosphorylated ERK (ppERK) as the monitored molecule, one can establish a time frame corresponding to its expected plateau concentration in healthy cells. If ppERK levels exceed this threshold prematurely, it may indicate aberrant signalling activity. This setup facilitates the evaluation of therapeutic interventions, such as the administration of dabrafenib or combination therapies, to determine their efficacy in restoring normal signalling dynamics or preventing the onset of dysregulation. This approach provides a valuable tool for simulating and assessing the impact of targeted treatments on cellular behaviour.

\begin{figure}[h]
\begin{center}
\frame{\includegraphics[width=\linewidth]{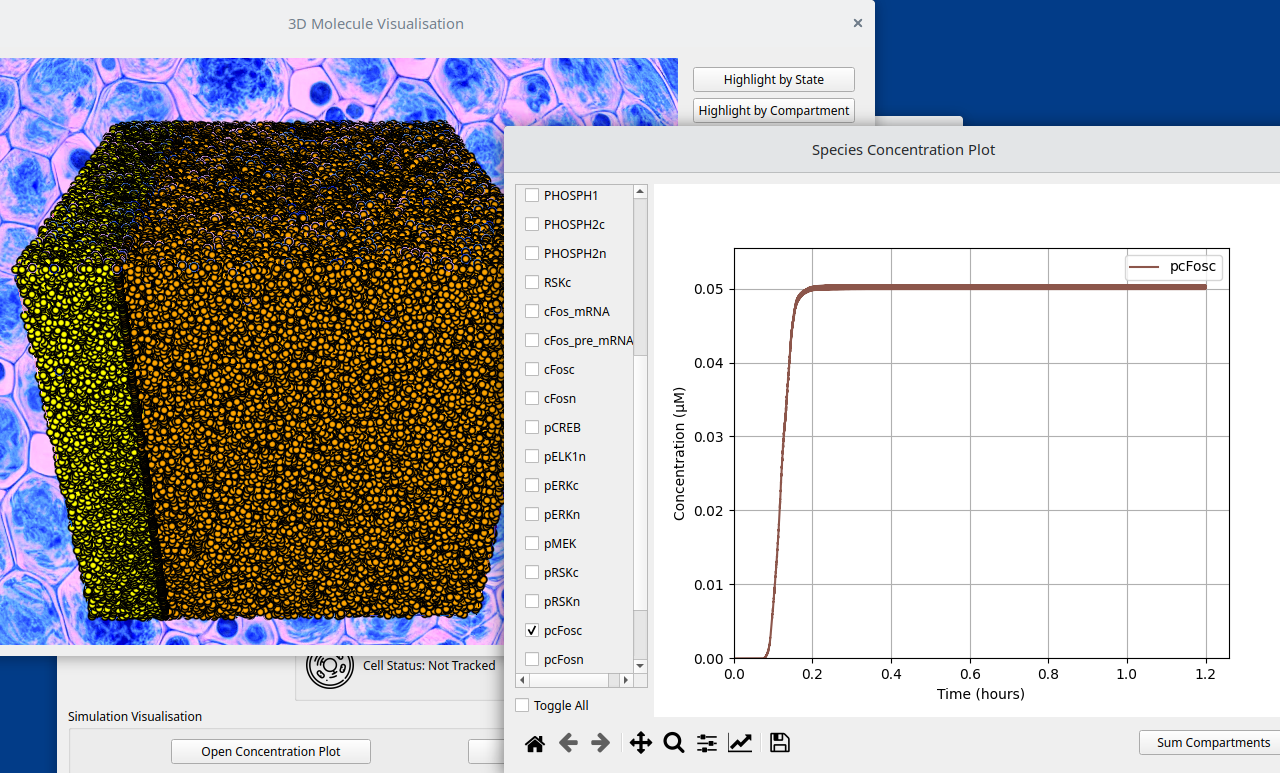}}
\caption{\textbf{The simulator interface during a running simulation.} The left panel displays the 3D visualisation of the simulation environment, highlighting molecules by compartment (orange for cytoplasm, yellow for nucleus). The right panel shows the real-time concentration plot of cytoplasmic phosphorylated cFos (pcFosc), a downstream product of the MAPK/ERK signalling pathway, reflecting the accumulation and phosphorylation of cFos until a steady-state level is reached. This example, based on Nakakuki et al.~\cite{nakakuki_ligand-specific_2010}, is discussed in Section~\ref{sec:emergent_dynamics} and demonstrates the simulator's ability to reproduce gene expression dynamics and compartmental processes. It also highlights the capability to perform simulations in a fully three-dimensional environment without requiring spatial slicing.}
\label{fig:gui_interface}
\end{center}
\end{figure}

\begin{figure}[h]
\centering
\frame{\includegraphics[width=\linewidth]{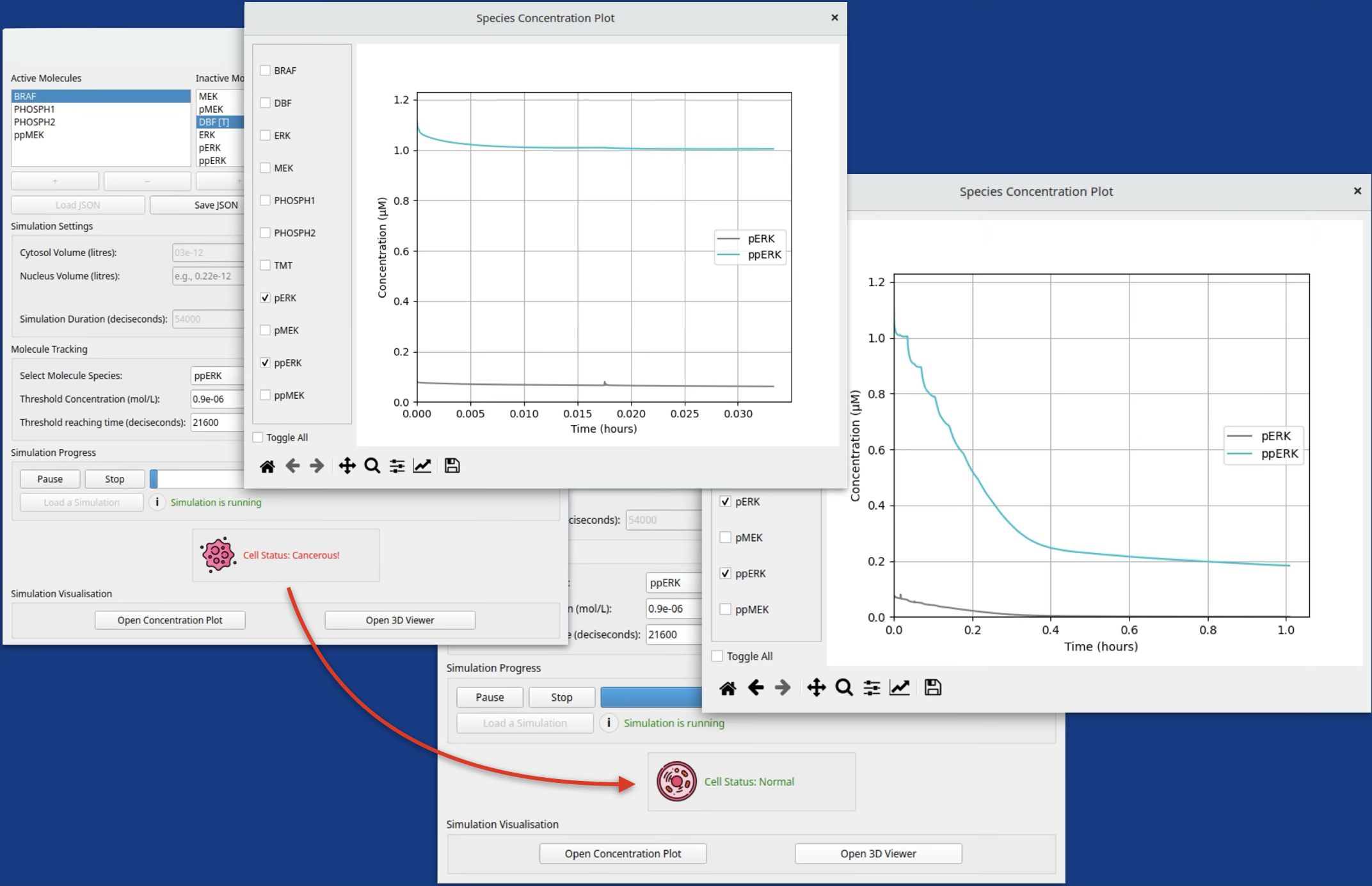}}
\caption{\textbf{Threshold-based monitoring of ppERK and treatment response in the simulator.}
Real-time classification of cell state from ppERK levels before and after dabrafenib. In the untreated baseline (left), the initial ppERK concentration is already above the healthy range, so the simulator classifies the cell as \textit{Cancerous}. With dabrafenib (right), ppERK declines to a level typical of normal cells and the status switches to \textit{Normal}. The plots show pERK (grey) and ppERK (cyan); the status icon updates automatically as ppERK crosses the preset threshold. This example corresponds to the dabrafenib case study reported in Section~\ref{sec:drug_response} of the Results.}
\label{fig:drug_effects}
\end{figure}

Simulation results are visualised through two tools available within the interface. A real-time plot displays the temporal evolution of molecule concentrations, while a three-dimensional viewer provides insights into spatial dynamics (see Fig.~\ref{fig:gui_interface}).
Combined, these tools support interactive exploration of treatment strategies by allowing users to observe and adjust, in real time, the effects of therapeutic interventions on molecular activity and disease-related cell states (see Fig.~\ref{fig:drug_effects}).

\section{Results}
\label{sec:results}

To evaluate the predictive power and biological fidelity of the GPU-based agent model, we designed two complementary case studies that reflect distinct aspects of MAPK/ERK signalling. 
The first focuses on the pharmacodynamic effect of the BRAF inhibitor dabrafenib on BRAFV600E melanoma, a setting that allows direct comparison against clinical trial data and previously published computational models. 
The second addresses the downstream nuclear response to ERK activation by replicating the study of Nakakuki et al.~\cite{nakakuki_ligand-specific_2010}, which investigated the dynamics of cFos expression and phosphorylation. 
Together, these case studies provide a stringent test of the simulator's ability to both reproduce dose--response profiles under targeted therapy and capture emergent pathway dynamics without significantly relying on predefined kinetic parameters.  

\subsection{BRAFV600E Inhibition: Dabrafenib Dose--Response Analysis}
\label{sec:drug_response}

We first assessed performance in the pharmacological context by comparing the model against (i) the deterministic ordinary differential equation (ODE) model of Hamis et al.~\cite{hamis_quantifying_2021} and (ii) our earlier agent-based prototype implemented on Repast4Py.  
To ensure a faithful comparison across the different simulation methods, we reproduced the dabrafenib-induced reduction in phosphorylated ERK---used as a downstream readout of BRAFV600E inhibition---across the same eight dabrafenib administrations considered in Falchook et al.'s clinical trials and also tested in our previous work: 138.5, 179, 205, 404, 560, 572.5, 726, and 908 ng/mL~\cite{maestri_cutting_2025}. In addition to these eight concentrations, we included a control condition without dabrafenib (0~ng/mL), to represent the untreated system, and an upper bound at $1.9246\times 10^{-6}\,\mathrm{M}$ (1000~ng/mL), corresponding to the maximum exposure reported in Hamis et al.~\cite{hamis_quantifying_2021}.

Each simulation scenario replicated eight hours of biological time---starting from the same initial setup described in our CIBB 2024 article and in Table~\ref{tab:SPECIES}---and, thanks to the GPU-accelerated FLAME GPU 2 implementation, all runs were completed in approximately the same wall-clock time (when no volume scaling is applied to enhance performance; see Section~\ref{sec:performances}). This represents a significant advancement over our previous CPU-bound implementation, demonstrating the new model's ability to maintain high temporal resolution while preserving execution speed.

\begin{table}[htbp!] \small
\caption{\textbf{Molecular species represented in the dabrafenib FLAME GPU 2 ABM and their initial concentrations.}
Note that the BRAF initial concentration matches that used in the model of Hamis et~al.~\cite{hamis_quantifying_2021} for non-drug-resistant patients (3~nM) and differs from the value adopted in the Repast4Py ABM, where it was set to $1\times10^{-8}$~mol/L (10~nM) to compensate for the artificial dilution introduced by the \textit{slicing} of the simulation volume. Refer to~\cite{hamis_quantifying_2021} and~\cite{maestri_cutting_2025} for further details on the model setup.}

\label{tab:SPECIES}
\centering
        \begin{tabular}{l@{\hspace{2em}}l}
        \toprule
         \textbf{Species}  & \textbf{Initial concentration (mol/L)} \\
        \midrule
         BRAF  & $\mathbf{3\times10^{-9}}$ \\
        dabrafenib & various concentrations (see text)\\
         ERK & 0 \\
        pERK & $8.0074\times10^{-8}$ \\
         ppERK & $1.1185\times10^{-6}$ \\
        MEK & $1.2\times10^{-6}$ \\
         pMEK & 0 \\
        ppMEK &  0 \\
         phosphatase 1 & $3\times10^{-10}$ \\
        phosphatase 2 & $1.2\times10^{-7}$ \\
        \bottomrule
    \end{tabular}    
    
\end{table}

\subsection{Simulation Performance Factors}
\label{sec:performances}

All simulations were initially executed on a Google Cloud Virtual Machine equipped with 4 vCPUs, 16 GB RAM, and an NVIDIA L4 GPU. Successive multiple-seed tests were conducted on the CINECA Leonardo (Booster partition) using one node equipped with one NVIDIA A100 GPU and 8 CPU cores (Intel Xeon Platinum 8358 CPU, 2.6 GHz ``Ice Lake''), as well as 512 GB of DDR4 memory. 

Despite the substantially higher peak performance of Leonardo relative to the Google Cloud VM, the simulation runtime per seed did not improve markedly. This is consistent with the workload being memory-bound and synchronisation-dominated rather than compute-bound. Step-wise host--device synchronisations, irregular (agent-based) memory access, and periodic CSV output scale weakly with GPU FLOPs; therefore, the A100’s extra cores and tensor throughput offer limited benefit for our kernels. In practice, the advantage of Leonardo was its throughput (the capacity to run many seeds concurrently) rather than per-seed latency.

In contrast with the hardware comparison above, runtime is chiefly governed by the simulated volume. At fixed concentrations, enlarging the domain increases the number of molecule-agents in direct proportion to volume. Because the per-agent neighbourhood work remains similar, the per-step cost (and hence wall-clock time) scales near-linearly with the agent count, i.e., approximately linearly with volume. This volume-driven behaviour, rather than raw GPU peak FLOPs, is the primary determinant of per-seed latency.

For the MAPK/ERK case study, we down-scaled the cytoplasmic domain used for multi-seed runs from the nominal $3\times 10^{-12}\,\mathrm{L}$ (approximately 14.4~\textmu m edge) to $3\times 10^{-13}\,\mathrm{L}$ (approximately 6.7~\textmu m edge). The tenfold reduction cut the wall-clock time required to simulate 8~h of biological time from ${\sim}10$~h to ${\sim}1.5$~h, enabling comprehensive multi-seed analysis within practical time and budget constraints. 

\begin{figure}[ht]
\centering
\includegraphics[width=\linewidth]{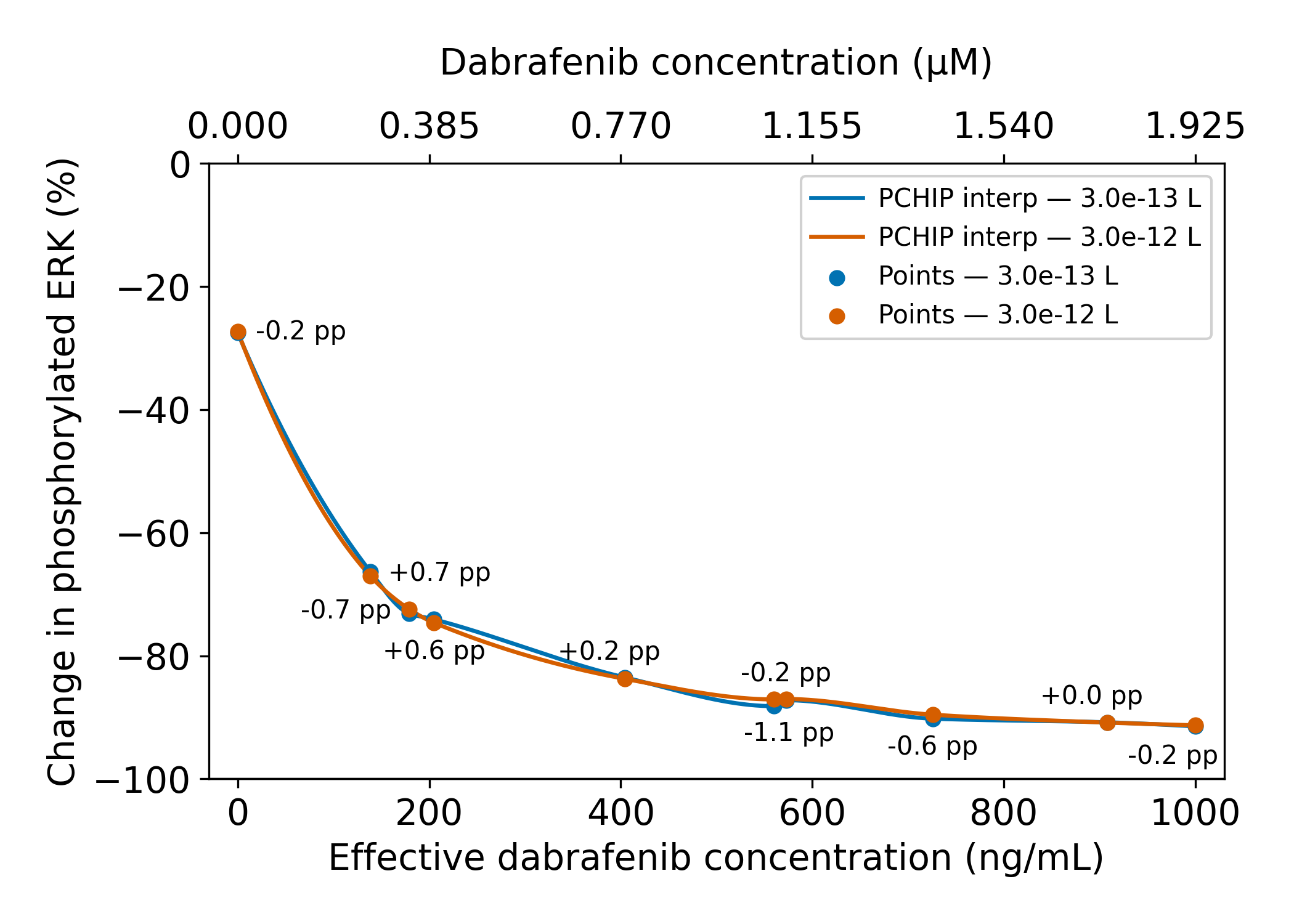}
\caption{\textbf{Comparison of different simulation volume runs.} 
Dose--response curves for the reduction of phosphorylated ERK (pERK+ppERK) are shown for cytoplasmic volumes of $3\times 10^{-12}\,\mathrm{L}$ and $3\times 10^{-13}\,\mathrm{L}$. 
Both curves were generated with the same seed (12) and connected using PCHIP interpolation. 
The results highlight that, despite the tenfold difference in simulated volume and the corresponding significant runtime reduction, the percentage decrease in phosphorylated ERK is nearly identical across doses, differing by at most about one percentage point (pp).}
\label{fig:comparison_volumes}
\end{figure}

Crucially, the biological readout was volume-invariant: across the doses for which both volumes were executed, the final pERK+ppERK decrease agreed to within ${\sim}1$ percentage point (pp) and the PCHIP-interpolated dose--response curves were visually almost indistinguishable (see Fig.~\ref{fig:comparison_volumes}). This invariance is consistent with the fact that the readout is a normalised percentage change at fixed concentrations, so scaling the domain rescales agent counts without altering concentrations or interaction probabilities. We therefore report multi-seed statistics at $3\times 10^{-13}\,\mathrm{L}$ and retain $3\times 10^{-12}\,\mathrm{L}$ single-seed runs as a validation check that down-scaling does not affect the inferred dose--response.

\subsection{Overcoming the limitations of the \textit{slicing} process}
\label{sec:overcoming_slicing}

The Repast4Py implementation relied on \emph{slicing} the cytoplasmic volume to curb CPU usage. To keep low-abundance species---most critically BRAF---from disappearing after this process, we were forced to enlarge the simulated cellular volume to an almost unrealistic 300~pL ($3\times10^{-10}\,\mathrm{L}$), a volume typical only of oocytes or adipocytes. 

\begin{figure}[htbp]
\centering

\begin{subfigure}[t]{.9\linewidth}
  \centering
  \includegraphics[width=\linewidth]{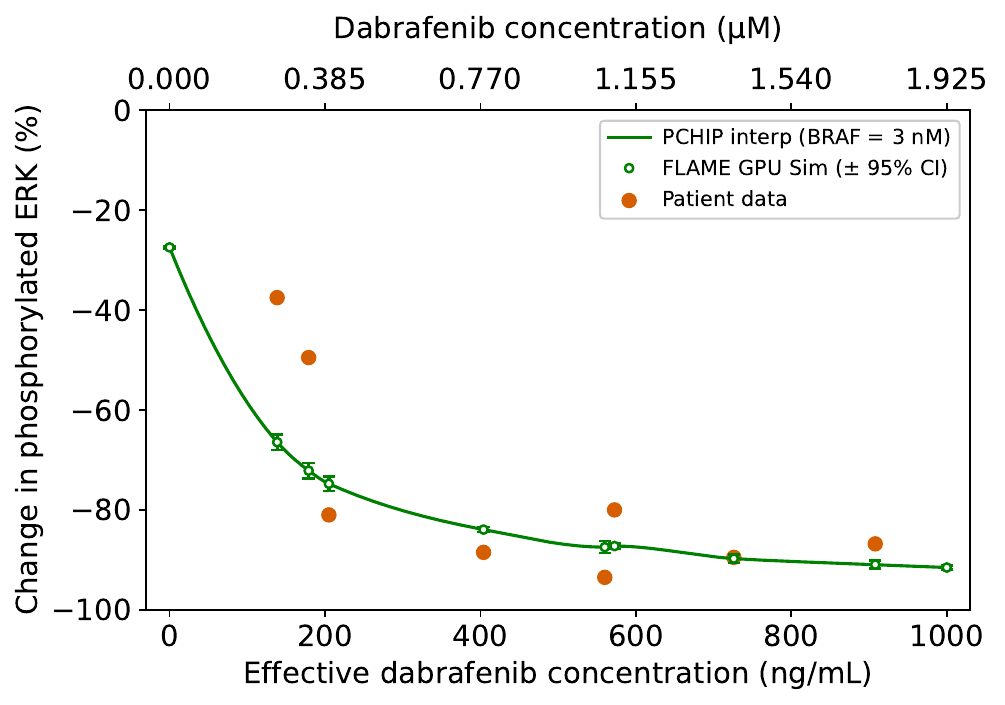}
  \caption{FLAME GPU~2 agent-based simulations}
  \label{fig:tri_flamegpu}
\end{subfigure}

\vspace{0.02\linewidth}

\makebox[\linewidth]{
  \begin{subfigure}[t]{0.49\linewidth}
    \centering
    \includegraphics[width=\linewidth]{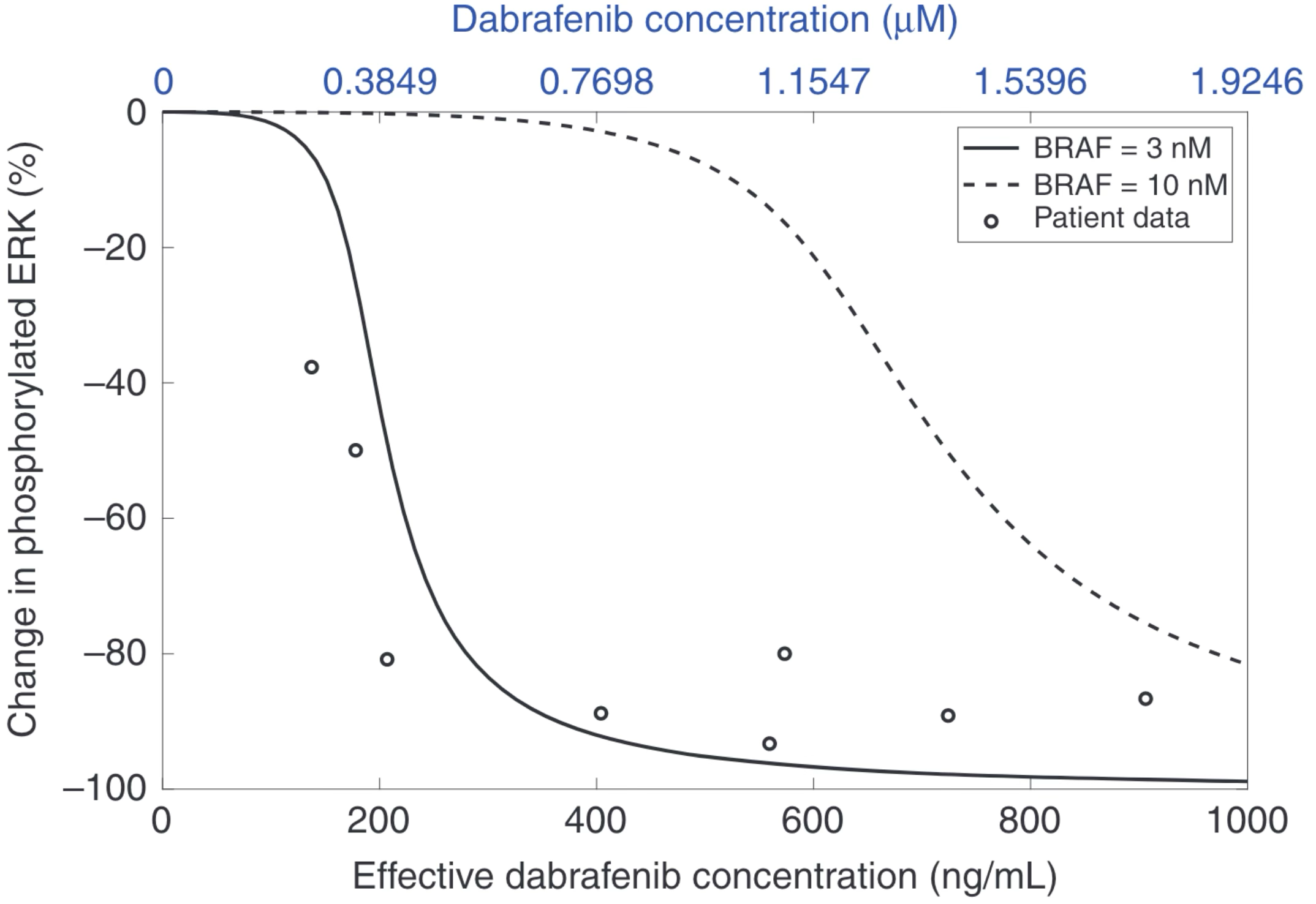}
    \caption{Hamis \emph{et al.} ODE simulations}
    \label{fig:tri_ode}
  \end{subfigure}
  \hfill
  \begin{subfigure}[t]{0.49\linewidth}
    \centering
    \includegraphics[width=\linewidth]{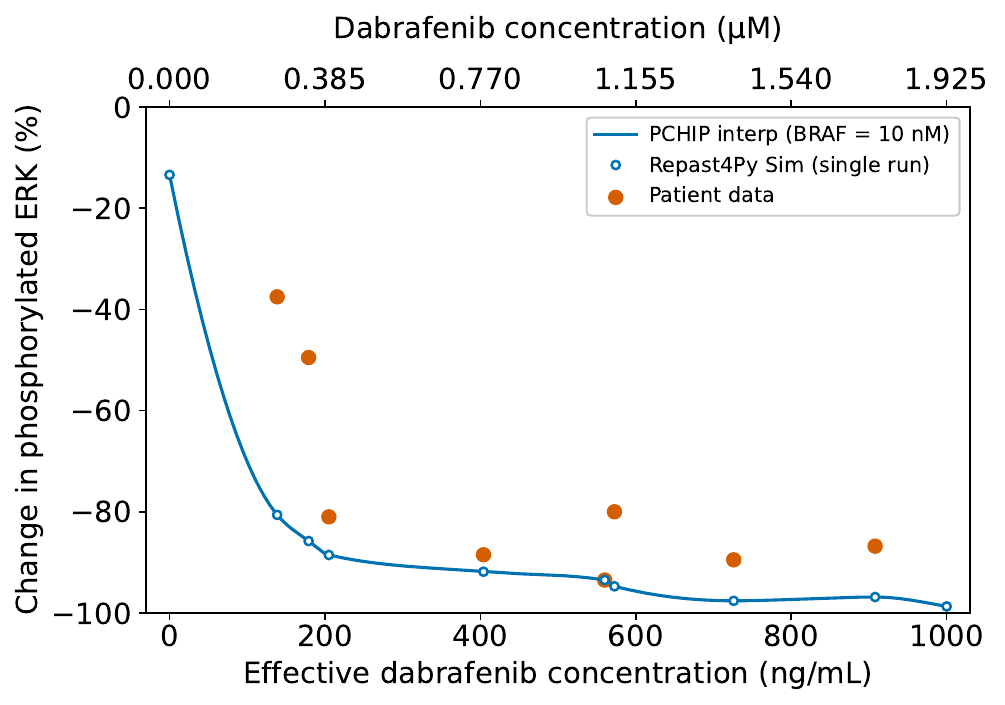}
    \caption{Repast4Py agent-based simulations}
    \label{fig:tri_repast}
  \end{subfigure}
}

\caption{\textbf{Dabrafenib dose--response at 8~h: reduction in phosphorylated ERK across different types of modelling and simulation.} 
\textbf{(a)} FLAME GPU~2 ABM (0.3~pL, unsliced): markers show the \emph{mean} of four stochastic seeds (12, 42, 314, 101010); vertical bars are 95\% confidence intervals (Student-$t$, $n{=}4$).
\textbf{(b)} ODE simulation adapted from Fig.~6 of Hamis et~al. (under CC BY 4.0 licence).
\textbf{(c)} Repast4Py ABM (300~pL, sliced).
In the two agent-based simulation panels, the line represents a \emph{shape-preserving PCHIP} interpolant through the reported dose points, chosen over cubic splines (as in~\cite{maestri_cutting_2025}) to avoid non-physical overshoot and to respect the expected monotonic dose--response. Circles denote patient data from Falchook et~al. The GPU ABM reproduces \textit{realistic cellular volumes} and \textit{concentrations of low-abundance species} (most critically, BRAF), without the extensive kinetic parameterisation required by the ODE model; moreover, it stays closest to the patient observations across both low and high doses.}
\label{fig:ppERK_decrease_comparison}
\end{figure}

Eliminating slicing and considering a fully 3D simulation environment allows us to simulate \textbf{3~pL} ($3\times10^{-12}\,\mathrm{L}$)---or \textbf{0.3~pL} ($3\times10^{-13}\,\mathrm{L}$) for multi-seed runs---while retaining the concentrations reported by Hamis et al.~\cite{hamis_quantifying_2021}. In particular, the model can now faithfully reproduce a system with \textbf{3~nM of BRAF}. In contrast, our previous Repast4Py-based prototype required increasing BRAF concentration to \textbf{10~nM} to match the experimental outcomes, a level associated by Falchook et al. with phosphorylated ERK changes typically observed only in drug-resistant patient-derived xenografts~\cite{falchook_dose_2014,xue_approach_2017} (see Fig.~\ref{fig:ppERK_decrease_comparison}).

\begin{figure}[h]
\vspace{3mm}
\begin{center}
\includegraphics[width=\linewidth]{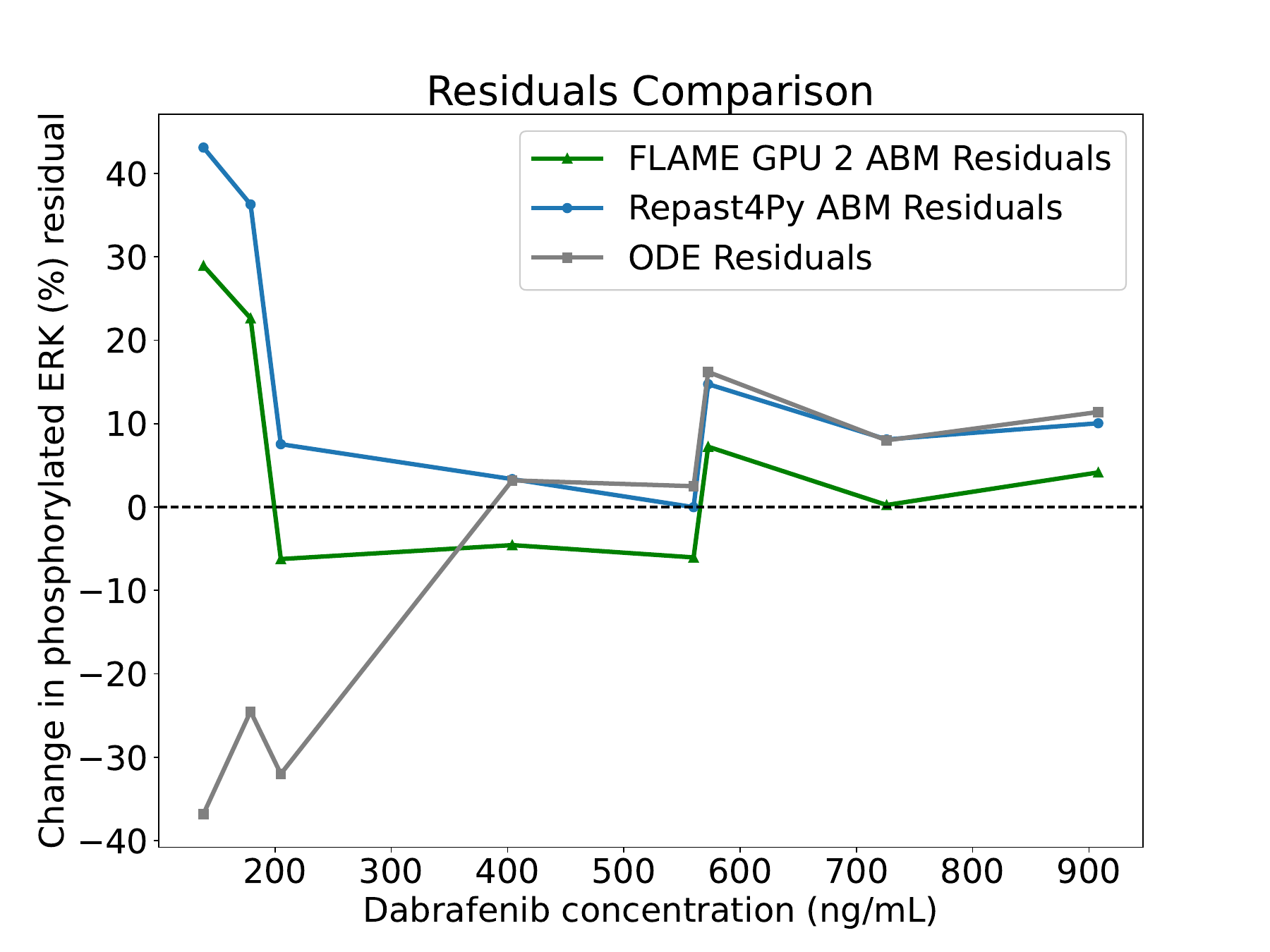}
\caption{\textbf{Residual analysis of ppERK reduction predictions across dabrafenib doses.}
The plot shows the percentage residual---model prediction minus patient observation---for phosphorylated ERK reduction at the eight dabrafenib exposure levels reported by Falchook et al.~\cite{falchook_dose_2014}. The light-grey squares represent the deterministic ODE model by Hamis et al., while the blue circles indicate the Repast4Py agent-based model with spatial slicing. The green triangles highlight the new GPU-accelerated agent-based model, which runs in a full 3D (unsliced) volume. While all models show variability, the GPU-based simulation generally stays closer to the zero line, particularly at low and high dabrafenib doses, suggesting improved consistency enabled by eliminating volume slicing and leveraging full 3D spatial computation.}
\label{fig:residuals}
\end{center}
\vspace{-8mm}
\end{figure}

As shown in Fig.~\ref{fig:residuals} and Table~\ref{tab:error_metrics}, the GPU-based model significantly improves predictive accuracy over both our Repast4Py-based prototype and the ODE model by Hamis et al., achieving the lowest error metrics across the board.

\begin{table}
      \centering
      \caption{\textbf{Comparison of prediction errors for ppERK reduction across different models.} The new GPU-based ABM, which operates on an unsliced 0.3~pL volume, achieves lower Mean Absolute Error (MAE) and Root Mean Square Error (RMSE) than both the Repast4Py ABM (which required spatial slicing at 300~pL) and the ODE model by Hamis et al.~\cite{hamis_quantifying_2021}. MAE reflects the average absolute difference between predicted and observed values, while RMSE captures the magnitude of larger deviations. The improved performance of the GPU model indicates greater numerical precision and consistency, enabled by avoiding volume slicing.}
\label{tab:error_metrics}
      \begin{tabular}{l@{\hspace{2em}}l@{\hspace{2em}}l}
        \toprule
        \textbf{Model} & \textbf{MAE} & \textbf{RMSE} \\
        \midrule
        FLAME GPU 2 ABM & \textbf{10.01} & \textbf{13.77} \\
        Repast4Py ABM & 15.40 & 21.29 \\
        ODE model & 16.83 & 20.77 \\
        \bottomrule
      \end{tabular}
    \end{table}

These findings support the suitability of the GPU-accelerated simulator as a computational aid for exploring targeted cancer treatments in silico. However, while this represents a promising step towards higher-fidelity pathway simulation, further experimental validation is necessary before the simulator can be considered a full-fledged virtual wet-lab for predictive use in a clinical setting.

\subsection{ERK-to-Nucleus Signalling: Reproducing cFos Expression Dynamics}
\label{sec:emergent_dynamics}

The GPU-based agent model also addresses a key limitation identified in simulating nuclear processes of gene expression and inhibition, which is due to the scarcity and, oftentimes, unreliability of kinetic data available in the literature. Our agent-based approach overcomes this issue by treating such processes as emergent from the local interactions of molecule-agents, rather than depending on predefined kinetic rate constants.

\begin{figure}[H]
\centering
\frame{\includegraphics[width=\linewidth]{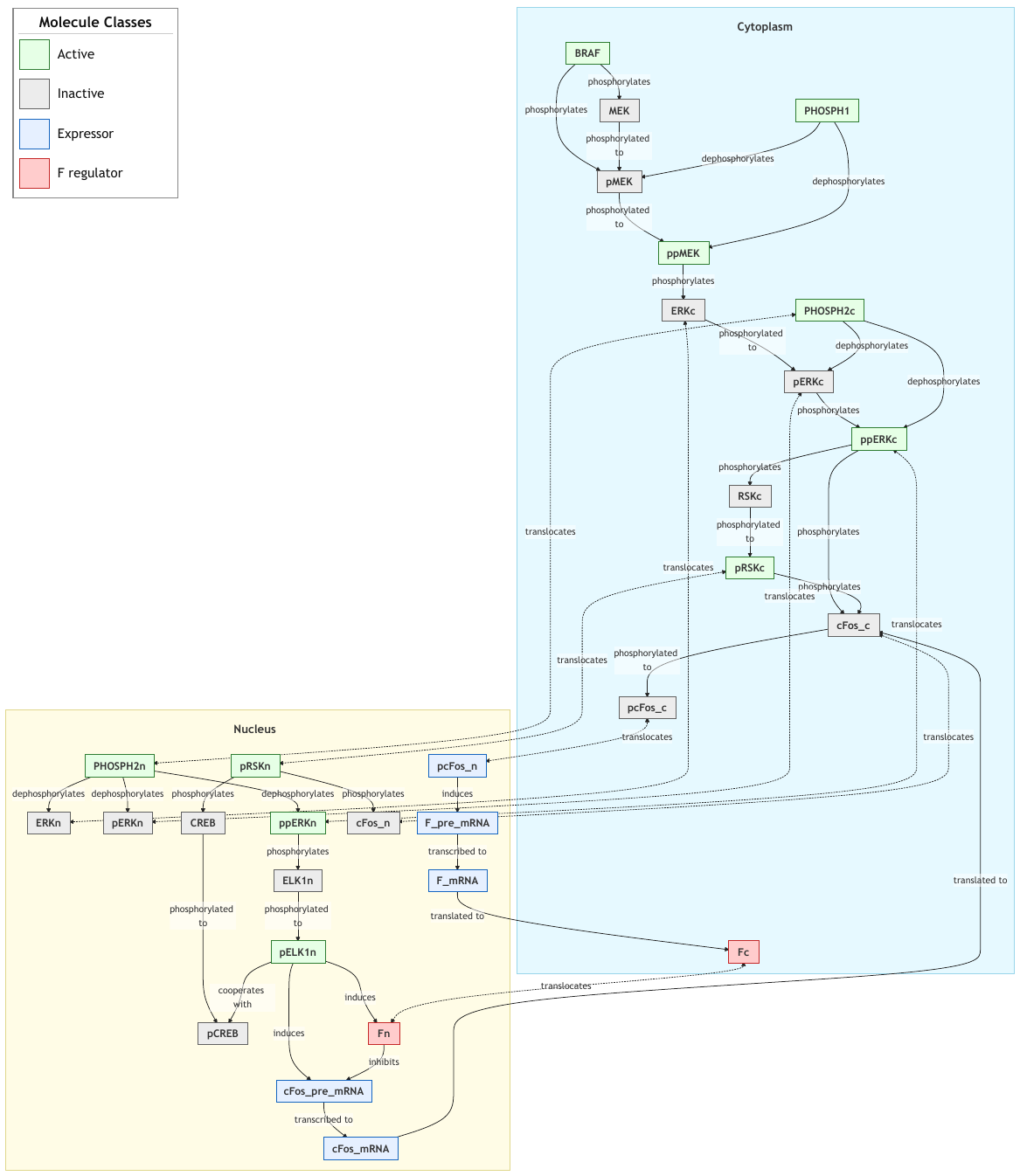}}
\caption{\textbf{Molecular interactions of the MAPK/ERK–cFos regulatory network as represented in the agent-based simulator.} The scheme distinguishes reactions occurring in the cytoplasm and in the nucleus, explicitly including nucleocytoplasmic translocation of intermediates (e.g., ERK, RSK, phosphatases). This compartmental organisation reflects the spatiotemporal regulation described in Fig.~1E of Nakakuki et al.~\cite{nakakuki_ligand-specific_2010}. In addition to the canonical ERK–RSK–Elk1/CREB–cFos cascade, the model incorporates a putative negative transcriptional regulator \textit{F} (highlighted in red and denoted \textit{Fn} in the nucleus and \textit{Fc} in the cytoplasm), hypothesised to be induced by pcFos and to inhibit cFos expression. Although the molecular identity of \textit{F} remains unknown, its inclusion enables the simulator to reproduce the transient mRNA dynamics observed experimentally. Together, these features highlight the ability of the GPU-based ABM to account for both compartmentalisation and transcriptional feedback in complex signalling systems.}
\label{fig:cfos_pathway}
\end{figure}

To demonstrate this capability, we replicated the study of Nakakuki et al.~\cite{nakakuki_ligand-specific_2010}, which investigated the expression of the transcription factor cFos and its phosphorylation downstream of MAPK/ERK activation. To reproduce the experimental conditions and pathway architecture described in that work, the simulator was configured to represent both cytoplasmic and nuclear compartments, allowing explicit modelling of protein translocation and transcriptional regulation. Molecules of ERK, once phosphorylated in the cytoplasm, translocate to the nucleus where they interact with transcriptional regulators such as Elk1 to induce the transcription of cFos pre-mRNA, its processing into mature mRNA, and subsequent translation into cFos protein. The cFos protein can then be further phosphorylated by activated ERK and RSK, yielding the phosphorylated cFos species (pcFos) that is experimentally measurable. This compartmentalised architecture is schematically illustrated in Fig.~\ref{fig:cfos_pathway}.

Moreover, consistent with the refinements introduced by Nakakuki and colleagues, the ABM includes an additional layer of transcriptional control beyond the canonical ERK–RSK–Elk1/CREB–cFos cascade. Their analysis showed that the experimental dynamics of cFos mRNA could not be explained solely by ERK kinetics, leading to the proposal of a negative feedback element acting at the transcriptional level. Our model adopts this mechanism by introducing an unknown factor, here denoted \textit{F}, which is hypothesised to be induced by pcFos and to repress, directly or indirectly, further transcription of cFos. Despite the molecular identity of \textit{F} remaining unresolved, the simulator is able to reproduce the experimentally observed transient wave of cFos mRNA and the downstream accumulation of pcFos. This demonstrates not only our framework's capability to match known pathway dynamics but also that it can accommodate and test hypotheses concerning regulatory components that are not yet fully characterised.

\begin{figure}[H]
\centering
\begin{subfigure}{0.9\linewidth}
  \centering
  \includegraphics[width=\linewidth]{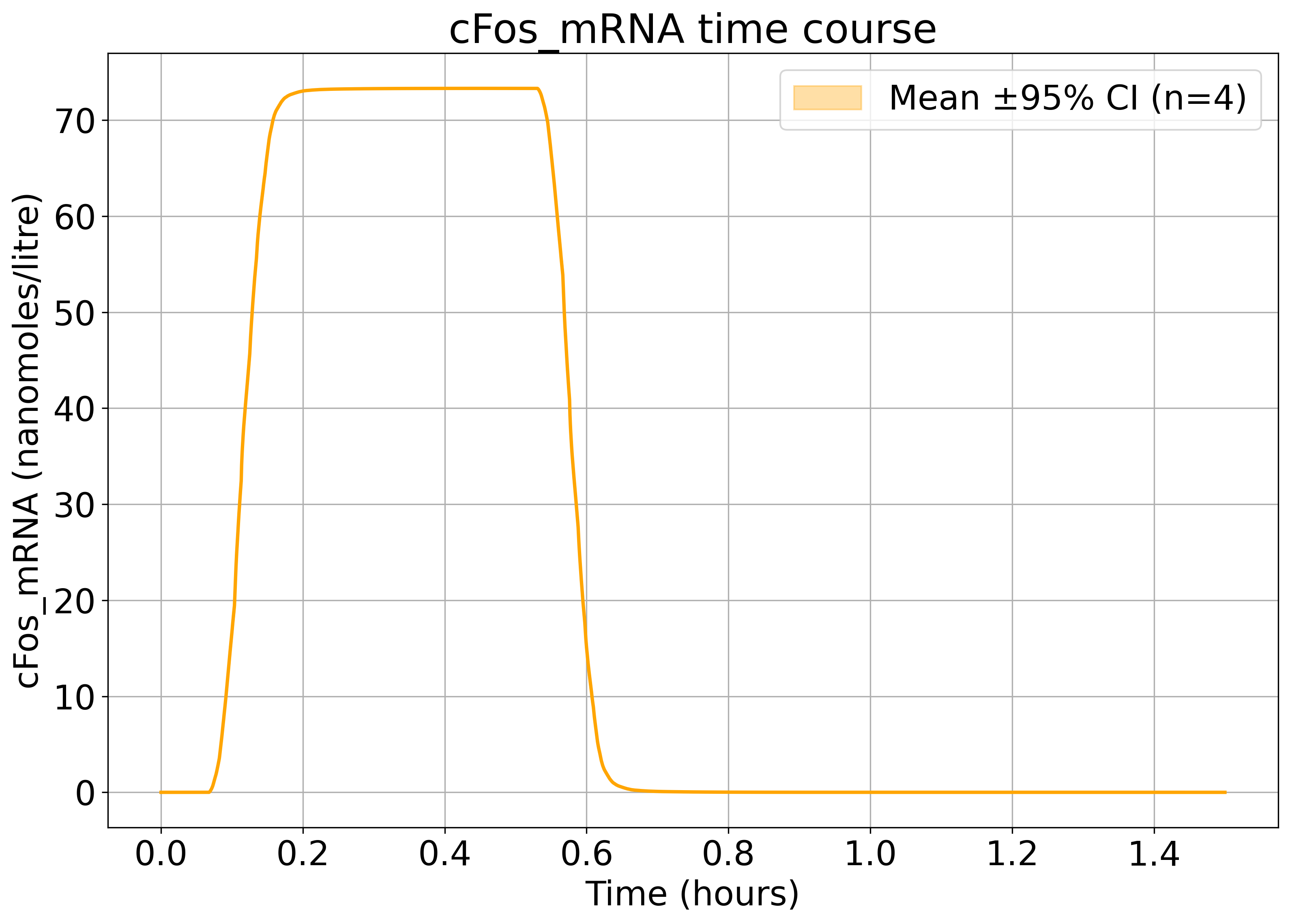}
  \caption*{\textbf{(a)}}
\end{subfigure}

\vspace{1em}

\begin{subfigure}{0.9\linewidth}
  \centering
  \includegraphics[width=\linewidth]{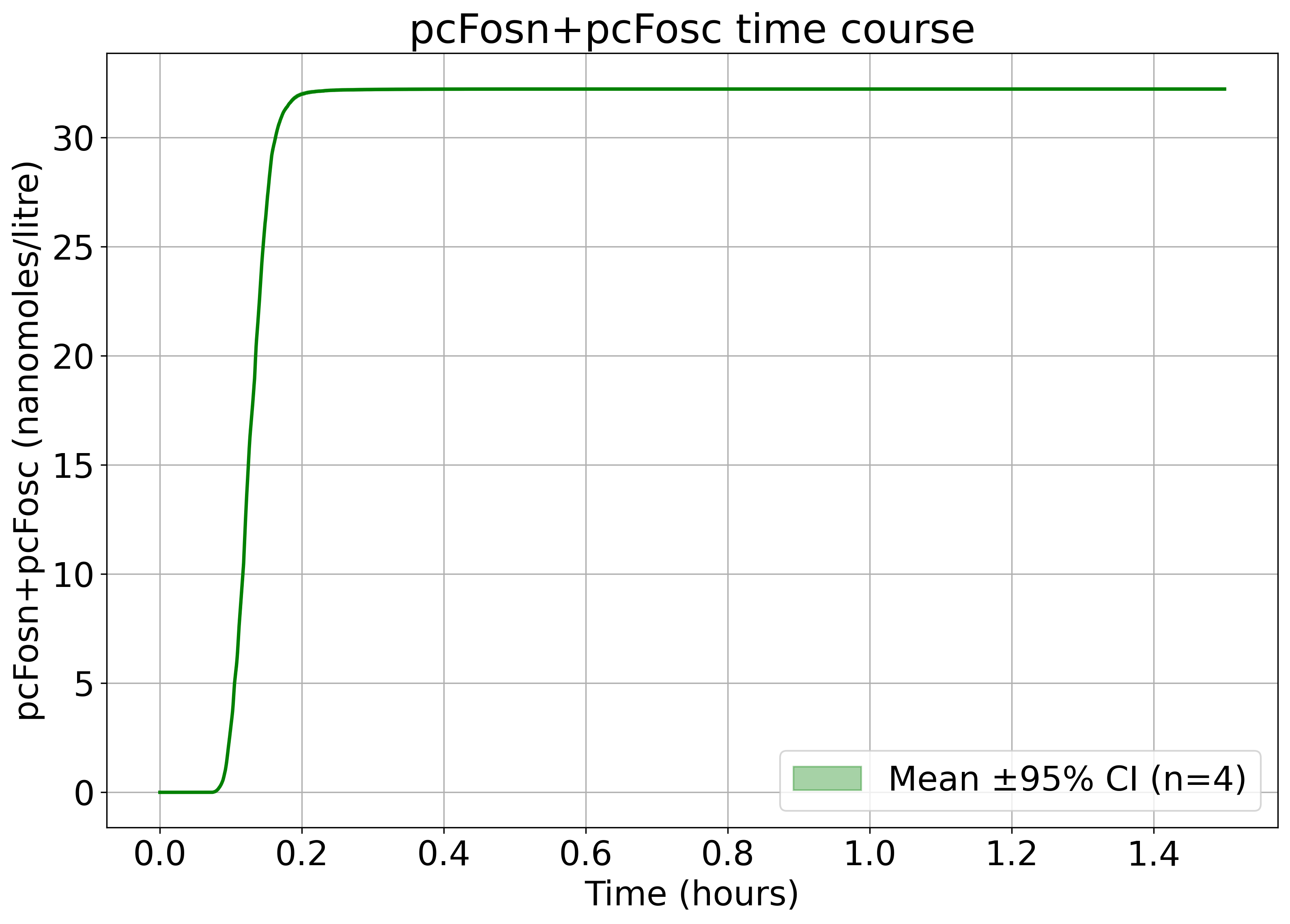}
  \caption*{\textbf{(b)}}
\end{subfigure}

\caption{
\textbf{Transient cFos mRNA and sigmoidal pcFos protein accumulation.} Panel (a) shows a transient peak in cFos mRNA concentration, typical of immediate-early gene expression subsequent to stimulation, followed by rapid degradation. Panel (b) depicts a sigmoid increase in phosphorylated cFos---nuclear (pcFosn) plus cytoplasmic (pcFosc)---representing gradual protein accumulation and activation until saturation. The simulations were run for a total duration of 90 minutes and replicated across four independent random seeds (12, 42, 314, and 101010; the same set used in the dabrafenib case study). As the trajectories from different seeds showed negligible variability, the confidence interval bands nearly coincide with the mean curves, explaining why the shaded regions are barely visible in the plots. These results are consistent with the known biological dynamics of the MAPK/ERK pathway, where short-lived transcriptional signals give rise to more stable protein responses.}
\label{fig:cfos}
\end{figure}

As demonstrated in Fig.~\ref{fig:cfos}, the GPU-based simulator successfully reproduced the experimental observations of Nakakuki et al., showing a transient peak of cFos mRNA followed by rapid degradation (Fig.~\ref{fig:cfos}a), and a sigmoidal accumulation of phosphorylated cFos protein (pcFos) over time (Fig.~\ref{fig:cfos}b). Importantly, these behaviours emerged naturally from the agent-based interactions and compartmental rules, without requiring explicit rate constants for molecular binding or dissociation. This highlights the ability of our approach to extend beyond cytoplasmic signalling cascades and reproduce nuclear responses and gene expression dynamics in a mechanistically faithful manner.

\section{Conclusion}
\label{sec:conclusions}

We have presented a GPU-accelerated agent-based simulator that addresses the key bottlenecks of our earlier CPU/HPC prototype, delivering biologically faithful, three-dimensional pathway simulations on a single workstation-class machine.
By eliminating the \emph{slicing} abstraction required by Repast4Py, the new implementation handles realistic cytoplasmic volumes (3~pL nominal; 0.3~pL for multi-seed runs) while maintaining experimentally reported concentrations, even for low-abundance species such as BRAF.

The evaluation across two case studies highlights complementary strengths of the approach. In the pharmacological context of BRAFV600E melanoma, the simulator reproduced the clinical dose--response of dabrafenib with higher predictive accuracy than both deterministic ODE formulations and a sliced ABM prototype, showing that it can provide robust pharmacodynamic predictions. In the second case study, focusing on the nuclear regulation of cFos, the model demonstrated its ability to capture compartmentalised dynamics and emergent gene expression responses. Importantly, it reproduced the transient transcriptional behaviour reported by Nakakuki et al. without requiring intensive kinetic parameterisation, and it accommodated the hypothesis of an unresolved negative regulator. These results emphasise the framework’s capacity not only to validate known pathway behaviour but also to test mechanistic hypotheses in cases where molecular identities remain unclear.

In this work, we focused on the MAPK/ERK signalling cascade. Still, we have already successfully extended our approach to simulate additional pathways, including PI3K/AKT, which will be presented in forthcoming works. 

Looking ahead, a natural development of the study will be to expand the repertoire of biological signals and treatments that can be faithfully simulated. This includes less-characterised pathways or those involved in understudied oncological conditions (e.g., simulating the inhibitory effect of targeted compounds on the PKA signalling pathway in fibrolamellar carcinoma).

Moreover, additional compartments (e.g., mitochondria, endoplasmic reticulum) will be introduced to investigate signalling that relies on organelle crosstalk and metabolite exchange.

These developments will further position our agent-based approach as a robust and high-resolution modelling strategy, suited to complement existing deterministic and stochastic frameworks for precision oncology research.

\begin{credits}

\subsubsection{\ackname} This study was funded by Diatech Pharmacogenetics S.R.L. through a research agreement with the University of Camerino (2022--2024).

\subsubsection{\discintname}
The simulations presented in this article were conducted using proprietary software developed as part of a company-funded project. The software and its source code cannot be shared publicly due to confidentiality agreements. The FLAME GPU 2.0.0-rc.2 framework on which the simulator is built is released under the \href{https://raw.githubusercontent.com/FLAMEGPU/FLAMEGPU2/v2.0.0-rc.2/LICENSE.md}{MIT licence}. Apart from this, the author has no competing interests to declare that are relevant to the content of this article.

\end{credits}

\printbibliography

\end{document}